\documentclass[aps,prd,twocolumn,superscriptaddress]{revtex4}
\usepackage{graphicx}
\usepackage{amsmath}
\usepackage{amsthm}
\usepackage{amsfonts}
\usepackage{amssymb}
\usepackage{dsfont}
\usepackage{multirow}
\usepackage{appendix}
\usepackage{slashed}
\usepackage[active]{srcltx}
\usepackage{psfrag}
\usepackage{subfigure}
\usepackage{capt-of}
\usepackage[utf8]{inputenc}
\begin{document}

\title{Modifications on the Properties of $D^*_{s0}(2317)$ as Four-quark State in Thermal Medium}
\date{\today}
\author{A. T\"{u}rkan}
\affiliation{\"{O}zye\u{g}in University, Department of Natural and
Mathematical Sciences, \c{C}ekmek\"{o}y,Istanbul,Turkey}
\author{J.Y.~S\"ung\"u}
\affiliation{Department of Physics, Kocaeli University, 41001
Izmit, Turkey}
\author{E. Veli Veliev}
\affiliation{Department of Physics, Kocaeli University, 41001
Izmit, Turkey}

\begin{abstract}

Since the low mass of $D^*_{s0}(2317)$ has still been a problem to
the conventional quark model, one can consider other options
regarding as multi-quark system. Therefore we investigate the
scalar open-charm state $D^*_{s0}(2317)$ by Thermal QCD Sum
Rules (TQCDSR) method using the two-point correlation function together
with contributions of the non-perturbative condensates up to
dimension six. Deriving and numerically analyzing thermal mass
and pole residue sum rules, we accomplish the effects on the
properties of $D^*_{s0}(2317)$ resonance in hot medium. Our
numerical evaluations indicate that the variations in mass and pole residue values are stable through the growing temperature up to
$T\cong100~\mathrm{MeV}$, but they begin to fall after this
point. At critical temperature, the values of mass
and pole residue change up to $91\%, 70\% $ of their values in vacuum in the molecular scenario and $ 91\%, 69\% $ in the diquark-antidiquark scenario. Our results does not give any definite information as to whether $D^*_{s0}(2317)$ resonance has molecular or diquark-antidiquark structure since they are very close to each other to differentiate them. Besides, we predict the hadronic parameters of the bottom partner of the $D^*_{s0}(2317)$ resonance in both molecular and diquark-antidiquark pictures. This bound state is worth investigating in future experiments. Also the detailed search of hot medium effects on the hadronic parameters
of open-charm meson $D^*_{s0}(2317)$ and the bottom partner could have some implications
to define the QCD phase diagram obtained from heavy-ion collision
experiments. Moreover these results can be useful in distinguishing
conventional quark model mesons from exotica.

\end{abstract}

\maketitle

%%%%%%%%%%%%%%%%%%%%%%%%%%%%%%%%%%%%%%%%%%%%%%%%%%%%%%%
\section{Introduction}
%%%%%%%%%%%%%%%%%%%%%%%%%%%%%%%%%%%%%%%%%%%%%%%%%%%%%%%

\begin{figure}[h!]
\centering
\includegraphics[height=8cm,width=8cm]{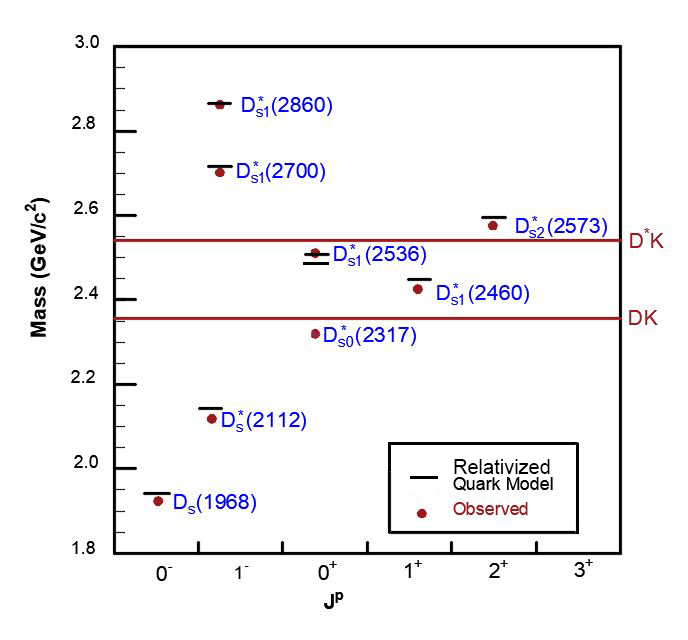}
\caption{Open-charm meson at $DK$ threshold \cite{Godfrey:1985xj,Zyla}.}	
\label{Fig}
\end{figure}

First seen in $D^*_{s0}(2317)\rightarrow D_{s}\pi^0$ by {\it
BABAR} $(2003)$ \cite{Aubert:2003fg}, $D_{s1}(2460)\rightarrow
D^*_{s}\pi^0$ by {\it CLEO} $(2003)$ \cite{Besson:2003cp} and
confirmed by {\it BELLE} $(2004)$\cite{Krokovny:2003zq}, a clear
experimental proof for the inner structure of $D^*_{s0}(2317)$ is
still unavailable \cite{Chen:2016spr}. The experiments observe a
narrow  mass below $DK$ threshold for the $D^*_{s0}$ state as
shown in Figure~\ref{Fig} while a number of models such as
Quark Model \cite{Godfrey:1985xj} and Lattice Theory find it above
close to $DK$ threshold \cite{Bali:2003jv,Lewis:2000sv}.
It also has very small width and only the upper
limit has been measured with the following experimental mass
value:
\begin{equation*}
M_{D_{s0}^\ast}(2317)=(2317.7 \pm 0.6)~\mathrm{MeV},~~
\Gamma_{D^*_{s0}} < 3.8~\mathrm{MeV}.
\end{equation*}
However, this width value is in disagreement with the Heavy Quark
Symmetry (HQS) estimation expecting to create a broad $1/2$ doublet with
$J^P=0^+,1^+$ \cite{GodfreyKokoski}.

The decay modes of $D_{s0}^\ast(2317)$ was reported respectively
by the {\it CLEO} and {\it BABAR} collaborations
\cite{Besson:2003cp, Aubert:2003pe} at the $90\%$ confidence
level:
\begin{eqnarray}\label{Eq:ratio1}
\frac{\Gamma(D_{s0}^*(2317) \to  D_{s}^\ast(2112) \gamma)}{
\Gamma(D_{s0}^*(2317) \to D_s\pi^0)} \left\{
  \begin{array}{ll}
    <0.052 & \hbox{{\it CLEO} \cite{Besson:2003cp}} \\
    <0.14 & \hbox{{\it BABAR} \cite{Aubert:2003pe}}
  \end{array}
\right.
\end{eqnarray}
and the branching ratio as follows:
\begin{eqnarray}\label{Eq:ratio3}
\frac{\mathcal{B}(D_{s0}^*(2317) \to  D_{s}^\ast(2112) \gamma)}{
\mathcal{B}(D_{s0}^*(2317) \to D_s\pi^0)} \left\{
  \begin{array}{lll}
    <0.059 & \hbox{{\it CLEO} \cite{Besson:2003cp}} \\
    <0.18 & \hbox{{\it BELLE} \cite{Abe:2003jk}}    \\
    <0.16 & \hbox{{\it BABAR}  \cite{Aubert:2006bk}}.\\
\end{array}
\right.
\end{eqnarray}
On the other hand, in $2018$ the absolute branching fraction
$\mathcal{B}(D_{s0}^*(2317)\rightarrow  D_s^{\pm}\pi^0$) is
measured as $1.00^{+0.00}_{-0.14}(stat)^{+0.00}_{-0.14}(syst)$
with a statistical significance of $5.8\sigma$ in the BESIII
detector at a center-of-mass energy $\sqrt{s}=4.6$ GeV for the
first time \cite{Ablikim:2017rrr}. It predicts that $D_{s0}(2317)$
should have a branching fraction of $\gamma D_s^{*-}$ at around
$15\%$ or even larger, but agrees well with the calculation in the
molecular picture \cite{Faessler:2007gv} which shows that the
branching fraction of $\pi^0 D_s^-$ is between $(93-100)\%$.

Flood of works are made on heavy-light, narrow open-charm systems
both in conventional and non-conventional frameworks such as using
Chiral Unitarity Model (CUM)\cite{Navarra:2015iea,Guo:2006rp},
Light-cone QCD Sum Rules (LCQCDSR)\ \cite{Colangelo:2005hv},
Effective Lagrangian Approach (ELA)\cite{Faessler:2007gv,
Xiao:2016hoa}, QCD Sum Rules (QCDSR)
\cite{Zhang:2018mnm,Wang:2015mxa,Wang:2006uba,Dai:2003yg,Colangelo:2003vg},
Lattice QCD (LQCD)\cite{Bali:2017pdv,Torres:2014vna},
Nonrelativistic Constituent Quark Model
(NRCQM)\cite{Ortega:2016aao}, etc.

Nevertheless many of these mesons in the open-charm sector can not be
well described by the Quark Model. This situation has opened a
discussion on their inner structures. The uncertainty with
conventional $c\bar{s}$ interpretation motivates many authors to
hypothesize that $D^*_{s0}(2317)$ state might be a molecule
\cite{Barnes:2003dj,Navarra:2015iea} or a diquark-antidiquark
state \cite{Cheng:2003kg,Nielsen:2005ia}. Further the existence of
a $ DK $ pole at this energy has been recently confirmed from Lattice calculations of scattering amplitudes. This state can be responsible of the bump near the
$ DK $ threshold around $ 2.4 $ GeV \cite{Cheung:2020mql}.

As the mass of $D^*_{s0}(2317)$ is about $40$ MeV below the
threshold of $DK$, it is thought to be most likely a $DK$ hadronic
molecule \cite{Barnes:2003dj}. They pointed out that such a
selection naturally manifests the anomalous mass of the
$D^*_{s0}(2317)$. In Ref.~\cite{Datta:2003re}, a model is proposed
to calculate the two body nonleptonic decays $B\rightarrow
D^{(*)}D_s (2317)(D_s(2460))$, presuming that the $D_s (2317)$ and
$D_s (2460)$ are $DK$ and $D^*K$ molecules. The coupled channel
prediction showed that the mass of $D^*_{s0}(2317)$ can conclude
from the strong coupling of the P-wave charmed-strange mesons to
the $DK$ \cite{Hwang:2004cd}. The decay behaviors of the
$D^*_{s0}(2317)$ was explored in the $DK$
 hadronic molecular picture \cite{Faessler:2007gv}. In
Ref. \cite{Bicudo:2004dx}, the $D^*_{s0}(2317)$ was considered as
kaonic molecule.  The studies in the Bethe-Salpeter approach
\cite{Xie:2010zza} and potential model pointed out that the
$D^*_{s0}(2317)$ could be a $DK$ hadronic molecule
\cite{Zhang:2006ix}.

Another possible explanation for the open-charm system
$D^*_{s0}(2317)$ is that it can be diquark-antidiquark state. The QCDSR
calculations also supported the idea that the $D^*_{s0}(2317)$
could be a diquark-antidiquark state \cite{Bali:2017pdv,Zhang:2018mnm}. In
Ref. \cite{Zhang:2018mnm} using four different possible currents
and including condensates up to dimension twelve found that
$D^*_{s0}(2317)$ state might be $0^+$ diquark-antidiquark state defined by scalar-scalar or axial-axial currents in QCDSR approach. Also, Bracco et al.
obtained an extremely narrow width $(\Gamma(D^{(1s)}_0\rightarrow
D_s\pi)=8~\mathrm{KeV})$ for $D^+_{s}(2317)$ in full QCD
\cite{Bracco:2005kt}.

Due to clarify this contradictory situation from a different point
of view, we investigate the $D_{s0}^\ast(2317)$ state in hot
medium. The variations in mass and decay constant of any hadron with temperature indicates that the QCD vacuum changes drastically. The
general opinion in the literature is that at extreme conditions
hadrons cannot survive as bound states, instead forming a new
state Quark-Gluon-Plasma (QGP)\cite{Ding:2015ona} and in this case
chiral symmetry is partially restored. The features of the
deconfined state of matter as well as the phase
boundaries from hadronic to quark-gluon degrees of freedom are yet discussed near
the so-called critical temperature $T_c\cong 155$ MeV
\cite{Bratkovskaya:2017gxq,Aoki:2006br,Andronic:2017pug}. Lattice QCD calculations indicate that the chiral symmetry restoration occurs at about the same
critical temperature  and energy density with
the deconfinement phase transition at vanishing baryon chemical
potential~\cite{Aoki:2006br,Andronic:2017pug,Steinbrecher:2018phh,Bazavov:2017dus}.

Howbeit there is no well-defined separation of phases by means of
the crossover nature of transition, thermal properties of hadrons
change expeditiously in the vicinity of $T_c$. As the temperature
raise, the quark condensate values are predicted to lessen from a
non-vanishing value in vacuum to $\langle q\bar{q} \rangle \approx 0$ which
coincides with chiral symmetry restoration. Therefore it is vital to find out experimental
observables which are sensitive to the quark condensates since $\langle q\bar{q} \rangle$
is not a measurable quantity. In this context,
to get clear signals from QGP many heavy-ion experiments will be
operated at the Nuclotron-based Ion Collider facility (NICA), future Facility for Antiproton and Ion Research (FAIR), as
well as RHIC at Large Hadron Collider (LHC).

Briefly, heavy-ion collision experiments at extreme conditions has vital importance to figure out the characteristics of hot hadronic matter, properties of QCD vacuum, chiral phase transitions and deconfinement. Though the hadrons produced in experiments has a very short lifetime, they interact with other particles in the fireball-medium where baryon density is expected to reach very high value and temperature comes close to the critical temperature. To explain these data precisely, the medium modifications of hadron parameters must be known. It is a main motivation to study the thermal behaviours of hadrons. Studying thermal effects on hadrons and their fate at extreme conditions may provide understanding about the collective behaviour of hot matter and also exploring the mechanisms behind the chiral symmetry breaking and confinement. 

In this paper, adopting the QCD Sum Rule approach to finite
temperature we evaluate the mass and  pole residue of
$D^*_{s0}(2317)$ treating it has four-quark content taking into
account quark, gluon and quark-gluon mixed condensates up to
dimension six assuming the quark-hadron duality is valid as well.
Note that the vacuum condensate expressions are displaced with
their thermal versions. This analysis can give us some hints on
the nature of the $D^*_{s0}(2317)$ and also may provide knowledge
about the systematics of strong interactions in hot medium.

This article is arranged as follows. In Section \ref{sec:MassPole}
we obtain the TQCDSR to evaluate the mass
and pole residue of $D^*_{s0}(2317)$. Section \ref{sec:Num} is
devoted to numerical analysis where we first present input
parameters used in the computation and give our numerical results for
the mass and pole residue of the considered state. These results
are confronted with the data obtained from other theoretical
models and available experimental data in the vacuum limit. Lastly, Appendix consists
of the explicit expressions of the obtained two-point thermal
spectral densities $\rho^{\mathrm{QCD}}(s, T)$ in the TQCDSR
theory.

%%%%%%%%%%%%%%%%%%%%%%%%%%%%%%%%%%%%%%%%%%%%%%%%%%%%%%%%%%%%%%%%%%%%%%%%%%%%
\section{Finite Temperature Sum Rules for the $D^*_{s0}(2317)$ State}
\label{sec:MassPole}
%%%%%%%%%%%%%%%%%%%%%%%%%%%%%%%%%%%%%%%%%%%%%%%%%%%%%%%%%%%%%%%%%%%%%%%%%%%

To explore the deviations of mass and pole residue of
$D^*_{s0}(2317)$ depending on increasing temperature, we extend
the QCDSR technique to TQCDSR. Calculation is initiated by writing
down the correlation function as in QCDSR model
\cite{Shifman,Reinders,Bochkarev:1985ex,Hatsuda:1992bv}:
\begin{equation}\label{eq:CorrF1}
\Pi(q,T)=i\int d^{4}x~e^{iq\cdot x}\langle\omega|
\mathcal{T}\{J(x)J^{\dag}(0)\}|\omega\rangle,
\end{equation}
where $\mathcal{T}$ represents the time ordering operator, $\omega $ denotes the hot medium, $T$ is
the temperature and $J(x)$ is the interpolating current
accompanying to $D^*_{s0}(2317)$ resonance. In thermal equilibrium,
thermal average of an operator $\mathcal{A}$ can be written as
follows:
\begin{equation}\label{eqn2}
\langle \mathcal{A}\rangle=Tr~(e^{-\beta \mathcal{H}}\mathcal{A})/Tr~(e^{-\beta \mathcal{H}}), \\
\end{equation}
here $\beta=1/T$ and $\mathcal{H}$ is the QCD Hamiltonian.

QCDSR technique which is very effective and practical can be used
to obtain the mass and  pole residue of the $D^*_{s0}(2317)$
state. In QCDSR, correlation function's dependence on momentum
provides us to extract the physical properties of any hadron by
calculating the correlator in two different momentum regions.The first is
called physical side which is described by hadronic observables
such as mass, residue, coupling constant of the considered
hadrons. The second is QCD side calculated in deep space-like region $ (q^2\ll\Lambda_{QCD}^2 ) $ in terms of QCD
parameters such as masses of quarks, quark-gluon condensates.
After getting correlation functions by converging these two
regions to each other, we equalize these two expressions under the assumption quark-hadron duality which sweeps the contributions of higher states under the carpet. 
%%%%%%%%%%%%%%%%%%%%%%%%%%%%%%%%%%%%%%%%%%%%%%%%%%%%%%%%%%%%%%%%%%%%%%%%%%%%
\subsection{{Physical Side}}
%%%%%%%%%%%%%%%%%%%%%%%%%%%%%%%%%%%%%%%%%%%%%%%%%%%%%%%%%%%%%%%%%%%%%%%%%%%
At low momentum ($q^2 \leq 0$ for $q^2$ space-like) or large
distance, in Eq.~(\ref{eq:CorrF1}) the interpolating current $J$
and its conjugate $J^{\dag}$ are interpreted as annihilation and
creation operators of the hadron. The correlation function is
saturated with a complete set of hadrons having the same quark
content and quantum numbers. This interpretation of the correlator
is called ``{\it Physical side}".

To extract the TQCDSR expressions we initially compute the
correlation function in connection with the physical degrees of
freedom. By integrating Eq.~(\ref{eq:CorrF1}) with respect to $x$, the
following equality is obtained (for brevity, we will use
$\mathcal{D}$ to represent $D^*_{s0}(2317)$ state in formulas):
\begin{equation}
\Pi^{\mathrm{Phys}}(q,T)=\frac{\langle
\omega|J|\mathcal{D}(q)\rangle\langle \mathcal{D}(q)|J^{\dagger
}|\omega\rangle}{m_{\mathcal{D}}^{2}(T)-q^{2}}+ \ldots,
\end{equation}
where $m_{\mathcal{D}}(T)$ is the temperature-dependent ground
state mass of scalar $D^*_{s0}(2317)$ particle and dots symbolize
contributions of the higher states and continuum which are
parametrized by means of the continuum threshold parameter $s_0$.
The definition of the temperature-dependent pole residue with the
matrix element is:
\begin{equation}\label{eq:Res1}
\langle \omega|J|\mathcal{D}(q)\rangle =\sqrt{2}f_{\mathcal{D}}(T)
m^4_{\mathcal{D}}(T).
\end{equation}
So the correlation function for the physical side can be written
with respect to the ground state mass and pole residue in the form
below:
\begin{equation}\label{eq:CorM}
\Pi^{\mathrm{Phys}}(q^2,T)=\frac{2 m_{\mathcal{D}}^{8}(T)f_{\mathcal{D}}^{2}(T)} {%
m_{\mathcal{D}}^{2}(T)-q^{2}}.
\end{equation}
After isolating the ground state contributions from the pole terms
by taking derivative, i.e applying Borel transformation, the
physical side is found as:
\begin{eqnarray}\label{eq:CorBor}
&&\mathcal{\widehat{B}}(q^{2})\Pi^{\mathrm{Phys}}(q^2,T)=2 m_{\mathcal{D}}^{8}(T)f_{\mathcal{D}}^{2}(T)~e^{-m_{\mathcal{D}}^{2}(T)/M^{2}},\quad
\end{eqnarray}
here $M$
is a sum rule parameter called as Borel parameter providing
elimination of the contributions from excited resonances and
continuum states.
%%%%%%%%%%%%%%%%%%%%%%%%%%%%%%%%%%%%%%%%%%%%%%%%%%%%%%%%%%%%%%%%%%%%%%%%%%%%
\subsection{{QCD Side}}
%%%%%%%%%%%%%%%%%%%%%%%%%%%%%%%%%%%%%%%%%%%%%%%%%%%%%%%%%%%%%%%%%%%%%%%%%%%

For high momentum ($q^2\geq0$), or short  distance, the
correlation function is evaluated with the help of Wilson's
Operator Product Expansion (OPE) due to the complex structure of
QCD vacuum. Employing OPE, contributions coming from quark, gluon
and mixed condensates can be included in the calculation of the
correlator and this second way of getting the correlation function
is called ``{\it QCD side}".

Now, our purpose is to define the QCD side in which the
correlation function is determined in terms of quark and gluon
degrees of freedom. In the first step of the calculation, selecting the
appropriate currents heavy and light quark fields are contracted and after some lengthy
calculations, the correlation function is obtained. 

Scalar molecular current defining $D^*_{s0}(2317)$ state can be
addressed in the following form~\cite{Albuquerque:2016nlw}:
\begin{equation}\label{curr_Scalar}
J(x)=\big(i\overline{d}^{a}\gamma_5
c^{a}\big)\big(i\overline{s}^{b}\gamma_5 d^{b}\big)
\end{equation}
where $a$ and $b$ are color indices.

Also $D^*_{s0}(2317)$ state can be regarded as diquark-antidiquark
and the current $J(x)$ can be defined by the following expression
\cite{Nielsen:2005ia}
\begin{eqnarray}\label{curr Scalar}
J(x)=\frac{\varepsilon^{ijk}\varepsilon^{mnk}}{\sqrt{2}}\big[\big(u^{T}_{i}C\gamma_5
c_{j}\big)\big(\overline{u}_{m}\gamma_5 C\overline{s}^{T}_n
\big)+u\rightarrow d\big],
\end{eqnarray}
here $i,j,k,m,n$ are color indexes, $\varepsilon$ is the
antisymmetric Levi-Civita tensor and $C$ is the charge conjugation matrix.

We can formulate the correlation function
$\Pi^{\mathrm{QCD}}(q^{2},T)$ as dispersion integral:
\begin{equation}
\Pi^{\mathrm{QCD}}(q^{2},T)=\int_{\mathcal{M}^2}^{\infty}\frac{\rho
^{\mathrm{QCD}}(s,T)}{s-q^{2}}ds+\widetilde{\Gamma}(q^{2},T),
\end{equation}
where $\mathcal{M}^2=(m_{c}+m_{s}+2m_{q})^2$ with $q=u$ or $d$
quark, $\rho^{\mathrm{QCD}}(s,T)$ is the spectral density  and $ \widetilde{\Gamma}(q^{2},T) $ symbolize the contributions coming from directly calculation of correlation function. The
spectral densities are calculated as an imaginary part of the
correlation function with the relation
$\rho^{\mathrm{QCD}}(s,T)=\frac{1}{\pi}Im[\Pi^{QCD}]$.

After some manipulations including the contraction of the quark fields, we obtain the QCD side of the correlation function in terms of the heavy and light quark propagators 
in the molecular and diquark-antidiquark scenarios, respectively:
\begin{eqnarray}
&&\Pi^{\mathrm{QCD}}(q^{2},T)=i\int d^{4}x~e^{iq \cdot x}\Big[Tr \Big(\gamma_5 S_d^{a'a}(-x) \gamma_5 S_c^{aa'}(x)\Big) \notag \\
&& \times Tr \Big( \gamma_5 S_s^{b'b}(-x)\gamma_5 S_d^{bb'}(x)\Big)\Big],
\end{eqnarray}
\begin{eqnarray}\label{eq:pi}
&&\Pi^{\mathrm{QCD}}(q^{2},T)=i A A' \int d^{4}x~e^{iq \cdot x}\Big[Tr \Big(\gamma_5  \widetilde{S}_u^{jj'}(x) \gamma_5\notag \\
&& \times S_c^{kk'}(x)\Big) Tr \Big(\gamma_5 \widetilde{S}_s^{nn'}(-x) \gamma_5 S_u^{m'm}(-x)\Big)+u\rightarrow d\Big], \notag \\
\end{eqnarray}
here 
\begin{eqnarray}
A=\frac{\varepsilon^{ijk}\varepsilon^{imn}}{\sqrt{2}},~~ A'=\frac{\varepsilon^{i'j'k'}\varepsilon^{i'm'n'}}{\sqrt{2}},	
\end{eqnarray}
and we employed short-hand notation $ \widetilde{S}^{jj'}(x) = CS^{jj'T}(x)C $ in Eq.~(\ref{eq:pi}).
Meanwhile, at finite temperature, due to failure of the Lorentz
invariance with the preferred reference frame and unveiling of the
residual $\mathcal{O}(3)$ symmetry, the additional operators
emerge in the short distance expansion of the product of two quark
bilinear operators and consequently, the thermal heavy and light
quark propagators contain new terms compared with the vacuum quark
propagators \cite{Mallik:1997pq}. Also we replaced the vacuum condensates by their thermal averages. The general definition of the
thermal heavy (charm) quark propagator $S_{c}^{ij}(x)$ can be
written as \cite{Reinders}:
\begin{eqnarray}\label{eq:HeavyProp}
S_{c}^{ij}(x)&=&i\int \frac{d^{4}k}{(2\pi )^{4}}e^{-ik\cdot x}\Bigg[ \frac{%
\delta _{ij}\Big( {\!\not\!{k}}+m_{c}\Big)
}{k^{2}-m_{c}^{2}}\nonumber \\
&-&\frac{gG_{ij}^{\alpha \beta }}{4}\frac{\sigma _{\alpha \beta }\Big( {%
\!\not\!{k}}+m_{c}\Big) +\Big(
{\!\not\!{k}}+m_{c}\Big)\sigma_{\alpha
\beta }}{(k^{2}-m_{c}^{2})^{2}}\nonumber \\
&+&\frac{g^{2}}{12}G_{\alpha \beta }^{A}G_{A}^{\alpha \beta
}\delta_{ij}m_{c}\frac{k^{2}+m_{c}{\!\not\!{k}}}{(k^{2}-m_{c}^{2})^{4}}+\ldots\Bigg],
\end{eqnarray}
here for the external gluon field $G_{ij}^{\alpha \beta}$ the
below short-hand notation is used;
\begin{equation*}
G_{ij}^{\alpha \beta }\equiv G_{A}^{\alpha \beta
}\lambda_{ij}^{A}/2,
\end{equation*}
where $\lambda_{A}^{ij}$ are the standard Gell-Mann matrices with
the number of gluon flavours $A=1,\,2\,\ldots 8$ being $i,\,j$ are
color indices. In Eq.~(\ref{eq:HeavyProp}) the first term gives
the perturbative contribution to the considered parameter while
the others, i.e nonperturbative terms contain gluonic additives.
In the nonperturbative terms the gluon field strength tensor
$G^A_{\alpha \beta } G_A^{\alpha \beta }(0)$ is fixed at $x = 0$
and the thermal light quark propagator $S_{q}^{ij}(x)$ is
expressed as \cite{Azizi:2014maa}:
\begin{eqnarray}\label{lightquarkpropagator}
S_{q}^{ij}(x) &=&i\frac{\slashed
x}{2\pi^{2}x^{4}}\delta_{ij}-\frac{
m_{q}}{4\pi^{2}x^{2}}\delta_{ij} -\frac{\langle \bar{q}q\rangle_T }{12}\delta_{ij} \nonumber \\&-&\frac{x^{2}}{192}%
m_{0}^{2}\langle \bar{q}q\rangle_T \Big[1-i\frac{m_{q}}{6}\slashed x \Big]%
\delta _{ij} \nonumber \\&+&\frac{i}{3}\Big[\slashed x
\Big(\frac{m_{q}}{16}\langle \bar{q}q\rangle_T
-\frac{1}{12}\langle u^{\mu} \Theta _{\mu \nu }^{f} u^{\nu}\rangle
\Big) \nonumber \\ &+&\frac{1}{3}(u\cdot x)\slashed u \langle
u^{\mu}\Theta _{\mu \nu }^{f} u^{\nu}\rangle
\Big]\delta _{ij}  \nonumber \\
&-&\frac{ig_{s}G _{ij}^{\mu \nu}}{32\pi ^{2}x^{2}}
\Big(\slashed x \sigma _{\mu \nu }+\sigma _{\mu \nu }\slashed
x\Big),\quad\quad
\end{eqnarray}
where $m_{q}$ indicates the light quark mass, $\langle
\bar{q}q\rangle_T $ is the light quark condensate as a function of
temperature, $m_0^2\equiv\langle 0 |\bar{q}g\sigma G q |0\rangle/\langle 0
\bar{q}q|0\rangle $ is extracted from sum rules of the nucleon channel ~\cite{Belyaev:1982sa} as the ratio of dimension-5 mixed quark-gluon condensate to dimension-3 chiral condensate, $u_{\mu }$ is the four-velocity of matter in hot
medium which is $u_\mu = (1, 0, 0, 0)$ in the matter rest frame
and $\Theta_{\mu \nu }^{f}$ is the fermionic part of the energy
momentum tensor. Additionally the following gluon condensate
expression depending on the gluonic part of the energy-momentum
tensor $\Theta _{\lambda \sigma }^{g}$ is used
\cite{Mallik:1997pq}:
\begin{eqnarray}\label{TrGG}
&&\langle Tr^{c}G_{\alpha \beta }G_{\mu \nu }\rangle
=\frac{1}{24}(g_{\alpha \mu }g_{\beta \nu }-g_{\alpha \nu
}g_{\beta \mu })\langle G_{\lambda \sigma
}^{a}G^{a\lambda \sigma }\rangle   \notag   \\
&&+\frac{1}{6}\Big[g_{\alpha \mu }g_{\beta \nu }-g_{\alpha \nu
}g_{\beta \mu }-2(u_{\alpha }u_{\mu }g_{\beta \nu }-u_{\alpha
}u_{\nu }g_{\beta \mu }\notag \\
&&-u_{\beta }u_{\mu }g_{\alpha \nu }+u_{\beta }u_{\nu }g_{\alpha \mu })\Big]%
\langle u^{\lambda }{\Theta }_{\lambda \sigma }^{g}u^{\sigma
}\rangle .
\end{eqnarray}
The next step to extract the thermal mass and pole residue sum
rules of the $D^*_{s0}(2317)$ state is to impose the Borel
transformation (i.e. omitting the continuum contribution taking
derivative) to the invariant amplitude
$\Pi^{\mathrm{QCD}}(q^{2},T)$ and selecting the same structures in
both physical and QCD sides, then equalizing the obtained
expressions with the related part of
$\mathcal{\widehat{B}}(q^{2})~\Pi^{\mathrm{Phys}}(q,T)$, lastly thermal pole
residue sum rule is found as:
\begin{equation}\label{eq:residueSR}
f_{\mathcal{D}}(T)=\sqrt{{n_1}/{d_1}}
\end{equation}
here,
\begin{equation}\label{eq:n1}
n_1=\int_{\mathcal{M}^2}^{s_{0}(T)}ds\rho^{\mathrm{QCD}}(s,T) e^{-s/M^{2}}+\mathcal{\widehat{B}}\mathrm{\widetilde{\Gamma}}(q^2,T),
\end{equation}
\begin{equation}\label{eq:d1}
d_1=2m_{\mathcal{D}}^{8}(T)e^{-m_{\mathcal{D}}^{2}(T)/M^{2}}.	
\end{equation}
To determine the thermal mass sum rule of $D^*_{s0}(2317)$ state, one must take the derivative of
Eq.~(\ref{eq:residueSR}) according to $1/M^2$ and so we get the
following analytic expression for the mass sum rule depending on temperature:
\begin{equation}\label{eq:massSR}
m_{\mathcal{D}}(T)=\sqrt{n_2/d_2}
\end{equation}
where,
\begin{equation}\label{eq:d2}
d_2=\int_{\mathcal{M}^2}^{s_{0}(T)}~ds~\rho^{\mathrm{QCD}}(s,T)~e^{-s/M^{2}}+\mathcal{\widehat{B}}\mathrm{\widetilde{\Gamma}}(q^2,T),
\end{equation}
\begin{eqnarray}\label{eq:n2}
&n_2&=\frac{d}{d(-1/M^2)}d_2.
\end{eqnarray}\\

In Eqs.~(\ref{eq:n1}) and (\ref{eq:d2}) $s_0(T)$ represent the thermal
continuum threshold parameter which is related with  $s_0(0)$ as introduced in Eq. (\ref{eq:sOT}) whose task is to separate the
contributions coming from the ground state and higher states. The two auxiliary parameters $ s_0 $ and Borel mass of the sum rule method will
be explained later in detail in Section\ \ref{sec:Num}.

Now we will perform numerical analysis to achieve to conclusion
considering $D^*_{s0}(2317)$ state according to both molecular
and diquark-antidiquark structure scenarios.

%%%%%%%%%%%%%%%%%%%%%%%%%%%%%%%%%%%%%%%%%%%%%%%%%%%%%%%%%%%%%%%%%%%%%%%%%%%%
\section{Numerical Calculation and Analysis}
\label{sec:Num}
%%%%%%%%%%%%%%%%%%%%%%%%%%%%%%%%%%%%%%%%%%%%%%%%%%%%%%%%%%%%%%%%%%%%%%%%%%%%

The TQCDSR calculations for the mass and  pole residue of
open-charm system  $D^*_{s0}(2317)$ involve some parameters e.g.
quark, gluon and mixed vacuum condensates and quark masses as
well. Values of these input parameters are compiled in the
Table~\ref{tab:input}.  
\begin{table}[htbp]
\caption{Input parameters.} \label{tab:input}
\begin{tabular}{cc}
\hline\hline
Parameters                               & Values \\
\hline\hline
$m_{u}$                                  & $2.16^{+0.49}_{-0.26}\mathrm{MeV} $ \cite{Zyla} \\
$m_{d}$                                  & $4.67^{+0.48}_{-0.17}\mathrm{MeV} $ \cite{Zyla} \\
$m_{s}$                                  & $93^{+11}_{-5}\mathrm{MeV} $ \cite{Zyla} \\
$m_{c}$                                  & $1.27 \pm 0.02~\mathrm{GeV} $ \cite{Zyla} \\
$\langle 0| \bar{q}q |0\rangle $         & $(-0.24\pm 0.01)^3~\mathrm{GeV}^3$ \cite{Shifman,Reinders} \\
$\langle  0| \bar{s}s  |0\rangle $       & $0.8\times\langle 0| \bar{q}q |0\rangle$ ~\cite{Shifman,Reinders} \\
$\langle\frac{\alpha_sG^2}{\pi}\rangle $ & $0.028(3)~\mathrm{GeV}^4$ ~\cite{Horsley:2012ra} \\
$m_0^2 $                                 & $(0.8\pm0.1) ~\mathrm{GeV}^2 $ \cite{Shifman,Reinders}  \\
\hline\hline
\end{tabular}
\end{table}

In addition to these parameters, to continue the calculations the
temperature-dependent quark, gluon condensates and also energy
density must be determined. As for the quark condensate, the fit
function provided from Ref.~\cite{Gubler:2018ctz}, which
intersects with the Lattice QCD data is utilized representing the
$u$ and $d$ quarks with $q$ :
\begin{eqnarray}\label{eq:qbarqT}
\frac{\langle\bar{q}q\rangle_{T}}{\langle 0| \bar{q}q |0\rangle}=
C_1 e^{aT}+C_2
\end{eqnarray}
and for the $ s $ quark
\begin{eqnarray}\label{eq:sbarsT}
\frac{\langle\bar{s}s\rangle_{T}}{\langle 0| \bar{s}s |0\rangle}=
C_3 e^{bT}+C_4,
\end{eqnarray}
here $ a=\mathrm{0.040~MeV^{-1}}$, $b=\mathrm{0.516~MeV^{-1}}$,
$C_1$$=-6.534\times10^{-4}$, $C_2=1.015$,
$C_3=-2.169\times10^{-5}$, $C_4$$=1.002$ are
coefficients of the fit function \cite{Azizi:2019cmj} and is valid up to a temperature $T=180~\mathrm{MeV}$ and $%
\langle 0|\bar{q}q|0\rangle $ represents the condensate of the
light quarks at vacuum.

The gluonic and fermionic pieces of the energy density can be
parametrized as in Ref.~\cite{Azizi:2019cmj} using the Lattice
QCD data presented in Ref.~\cite{Bazavov:2014pvz}:
\begin{eqnarray}\label{tetaf}
\langle u^\mu\mathrm{\theta}^{f}_{\mu\nu}u^\nu\rangle_T &=& (\xi_1
e^{cT}+C_5)~T^{4},
\end{eqnarray}
\begin{eqnarray}\label{tetag}
\langle u^\mu\mathrm{\theta}^{g}_{\mu\nu}u^\nu\rangle_T &=& (\xi_2
e^{dT}-C_6)~T^{4},
\end{eqnarray}
here $\xi_1=0.009 $, $c=24.876~\mathrm{GeV^{-1}}$,
$C_5=0.024$, $\xi_2=0.091 $, $d=21.277~\mathrm{GeV^{-1}}$
and $C_6=0.731$ \cite{Azizi:2019cmj}.

Also the temperature-dependent gluon condensate $ \langle
G^2\rangle_{T} $ is defined as in Ref. \cite{Gubler:2018ctz}:
\begin{eqnarray}\label{delta}
\delta \Big \langle \frac{\alpha_{s}G^{2}}{\pi}\Big
\rangle_{T}&=&-\frac{8}{9}[ \delta T^{\mu}_{\mu}(T)-m_{u} \delta
\langle\bar{u}u\rangle_{T}\nonumber
\\ &-&m_{d} \delta \langle\bar{d}d\rangle_{T}-m_{s} \delta
\langle\bar{s}s\rangle_{T}],
\end{eqnarray}
where the vacuum subtracted values of the considered quantities are
used as $ \delta f(T)\equiv f(T)-f(0) $ and $ \delta
T^{\mu}_{\mu}(T)=\varepsilon(T)-3p(T) $: $\varepsilon(T)$ is the
energy density and $ p(T) $ is the pressure. Taking into account
the recent Lattice calculations
\cite{Bazavov:2014pvz,Borsanyi:2013bia} we get the fit function of
$ \delta T^{\mu}_{\mu}(T) $ as \cite{Azizi:2019cmj}
\begin{eqnarray}\label{epsmines3p}
\frac{\delta T^{\mu}_{\mu}(T) }{T^{4}}&=&(C_7 e^{g T}+C_8)
\end{eqnarray}
with $C_7=0.020$, $h=29.412~\mathrm{GeV^{-1}}$,
$C_8=0.115$. 
Moreover we use the following expression for the  temperature-dependent strong coupling  \cite{Kaczmarek:2004gv,Morita:2007hv} 
\begin{eqnarray}\label{geks2T}
g_s^{-2}(T)=\frac{11}{8\pi^2}\ln\Big(\frac{2\pi T}{\Lambda_{\overline{MS}}}\Big)+\frac{51}{88\pi^2}\ln\Big[2\ln\Big(\frac{2\pi
T}{\Lambda_{\overline{MS}}}\Big)\Big]
\end{eqnarray}
where $\Lambda_{\overline{MS}}\simeq T_{c}/1.14$.

\begin{figure}[h!]
\centering
\includegraphics[width=8.2cm]{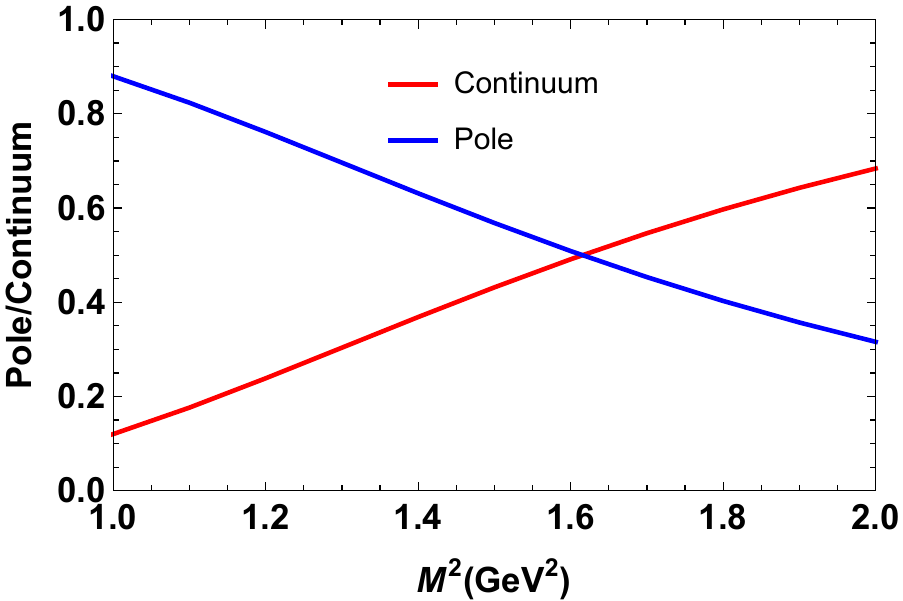}
\caption{The relative pole and continuum contributions of the $ J^P = 0^+ $ open charm state with $ \sqrt{s}=2.81 $ GeV at the diquark-antidiquark assumption as an example. } 
\label{PoleContTetra}
\end{figure}

Continuum threshold as a function of
temperature belonging to $D^*_{s0}(2317)$ state is another
auxiliary parameter that needs to be determined. The expression of
the continuum threshold in terms of temperature
is used as follows (for details see \cite{Dominguez:2016roi,Borsanyi:2010bp,Bhattacharya:2014ara}) :
\begin{eqnarray}\label{eq:sOT}
\frac{s_0(T)}{s_0(0)}= \bigg[ \frac{\langle \bar{q}q\rangle_T}{\langle 0| \bar{q}q |0 \rangle}\bigg]^{2/3}
\end{eqnarray}
where $s_0(T)$ is defined with $s_0(0)$ in vacuum threshold. It
is not random and relies on the mass of the first excited state of
the $D^*_{s0}(2317)$. That's why the chosen interval for the $s_0$
is to be relatively weak dependent from physical quantities for
the $D^*_{s0}(2317)$ state. As stated in the philosophy of the
QCDSR technique, the physical quantities shouldn't be connected
with the auxiliary parameters $M^2$ and $s_0$.  But in
real calculations these quantities nevertheless are sensitive to
the choice both of $M^2$ and $s_0$. Therefore, the parameters
$M^2$ and $s_0$ should be settled to minimize the dependence of
$m_\mathcal{D}$ and $f_\mathcal{D}$ on them. Convergence of the
OPE plus suppression of the contributions coming from the
higher states and continuum are a must for the sake of fixing the
working region of the Borel parameter $M^2$. As shown in Figure \ref{PoleContTetra}$, M^2_{max} $ should be $ 1.6~\mathrm{GeV}^2$. To determine the working region of continuum threshold we use $ \sqrt{s_0}=m_\mathcal{D} +(0.3;0.5) $ in the classical conjecture.\\

It is clear that $m_{\mathcal{D}}$ and $f_\mathcal{D}$ should not
depend on the the auxiliary parameters $M^2$ and $s_0$. The analysis carried
out by taking into account all of aforementioned constraints allow
us to fix the continuum threshold and   Borel parameter  as $1.2~\mathrm{GeV^2} \leq  M^2 \leq 1.6~\mathrm{GeV^2} $ and $  6.8~\mathrm{GeV^2} \leq s_0 \leq 7.9~\mathrm{GeV^2}$.

\begin{figure}[htbp]
\begin{center}
\includegraphics[width=8cm]{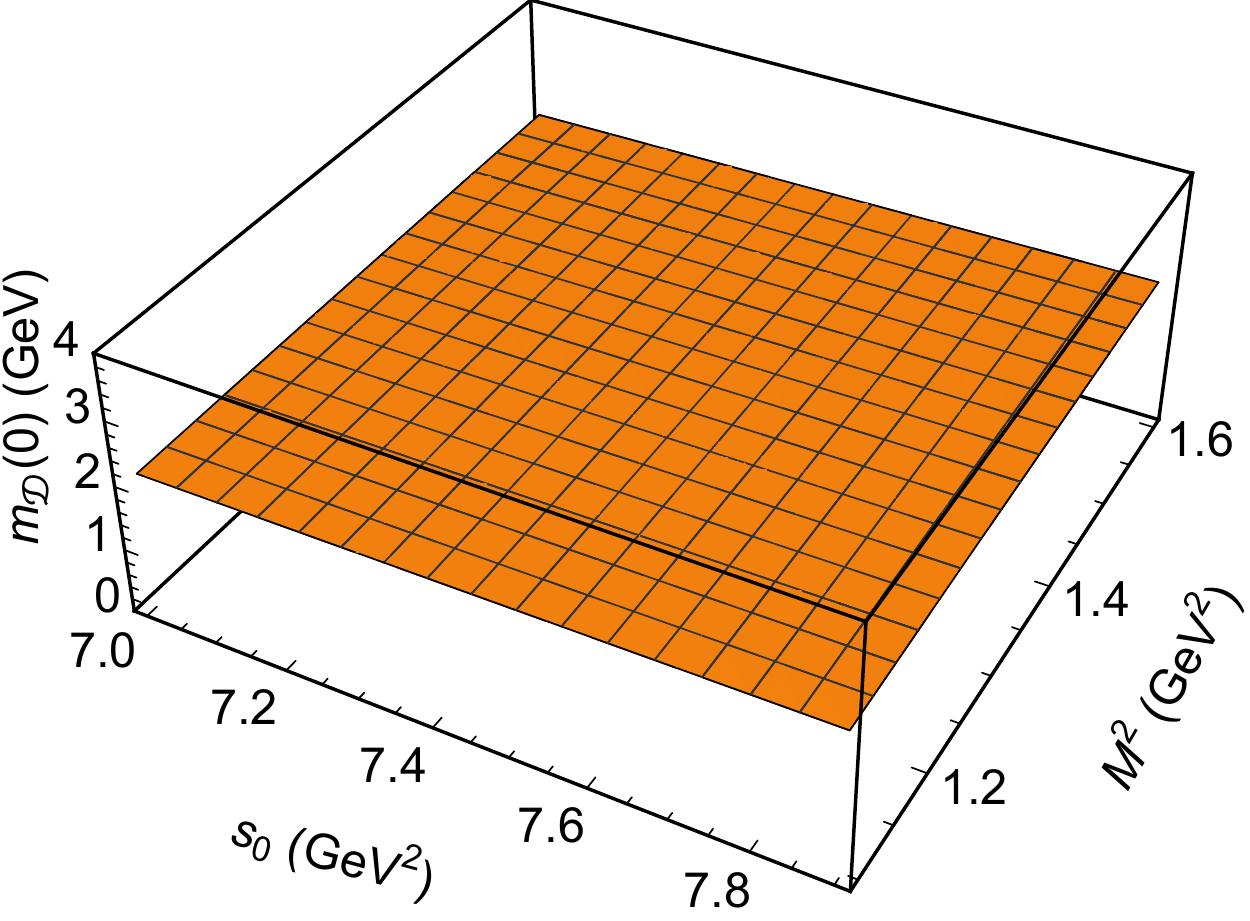}
\end{center}
\caption{$ 3D $ graph of the vacuum mass of the $D^*_{s0}(2317)$ state versus continuum threshold $s_0$ and Borel mass $M^2$ in the diquark-antidiquark picture.} \label{fig4}
\end{figure}

Note that the dependence of the mass and pole residue on $M^2$ is
stable in this interval, so we can trust that the obtained sum
rules will give accurate results. To show the independence of
physical quantities from $M^2$ and $s_0$, we present the $ 3D  $ plot of mass versus continuum threshold and Borel parameter $ M^2 $ in the diquark-antidiquark
picture in Figure \ref{fig4} and we see the stability of mass sum rule according to these parameters which is the main criterion of QCDSR.
\\

At $T=0$ open-charm system $D^*_{s0}(2317)$ is investigated with
many theoretical model both in the molecular and diquark-antidiquark
pictures \cite{Barnes:2003dj}-\cite{Bracco:2005kt}. Our results in $ T=0 $ limit TQCDSR model are presented below both in molecular and diquark-antidiquark scenarios: \\
\begin{eqnarray*}  \label{eq:OurResults}
m_{\mathcal{D}}^{mol.}&=&2316^{+33}_{-34}~\mathrm{MeV},~~ f_{\mathcal{D}}^{mol.}= 82.1^{+2.1}_{-2.1}~\mathrm{keV},\\\\
m_{\mathcal{D}}^{di.}&=&2317^{+33}_{-34}~\mathrm{MeV},~~~ f_{\mathcal{D}}^{di.}=93.1^{+2.3}_{-2.4}~\mathrm{keV}.
\end{eqnarray*}

\begin{figure}[htbp]
\begin{center}
\includegraphics[width=8.5cm]{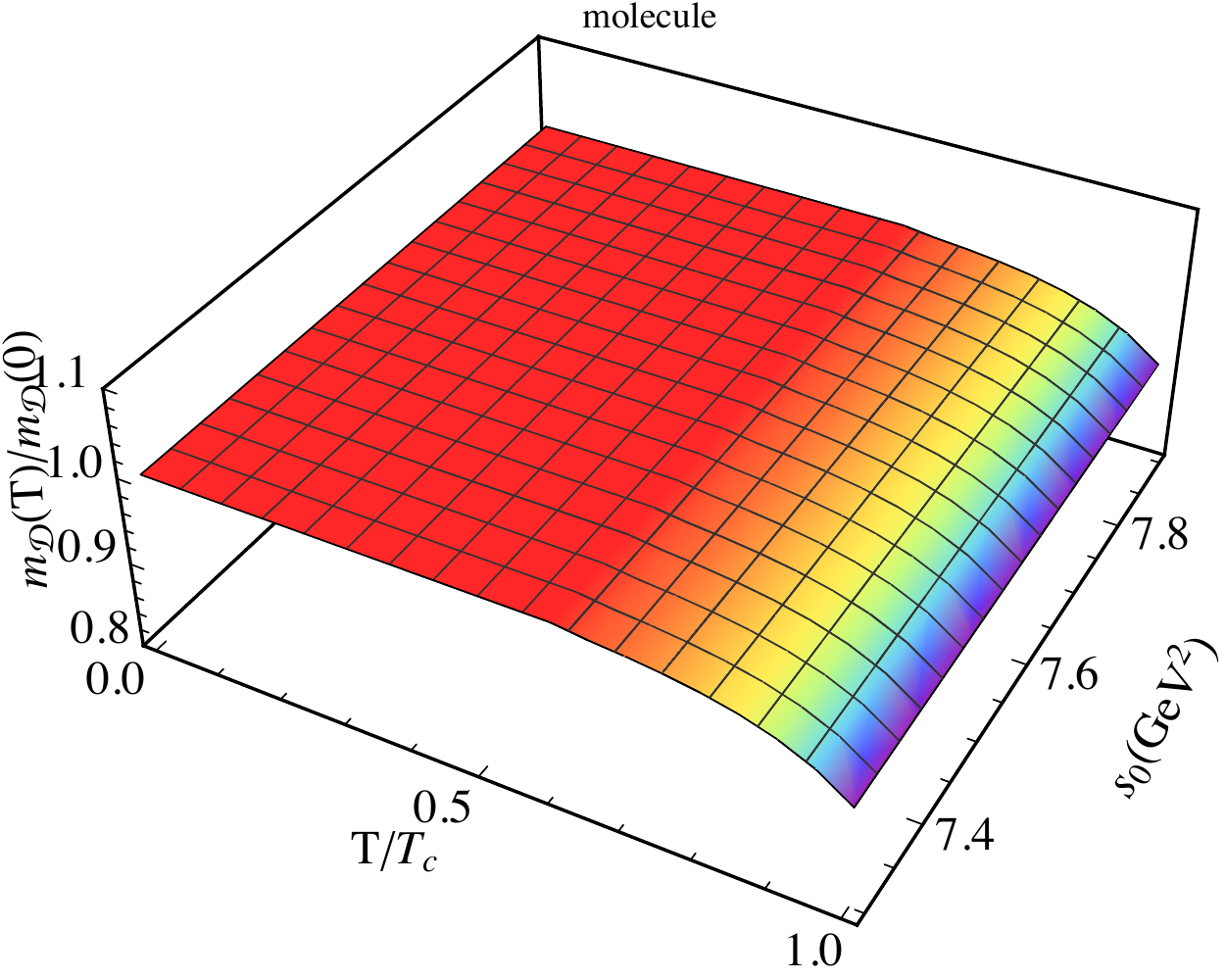}\\[0.7cm]
\includegraphics[width=8.5cm]{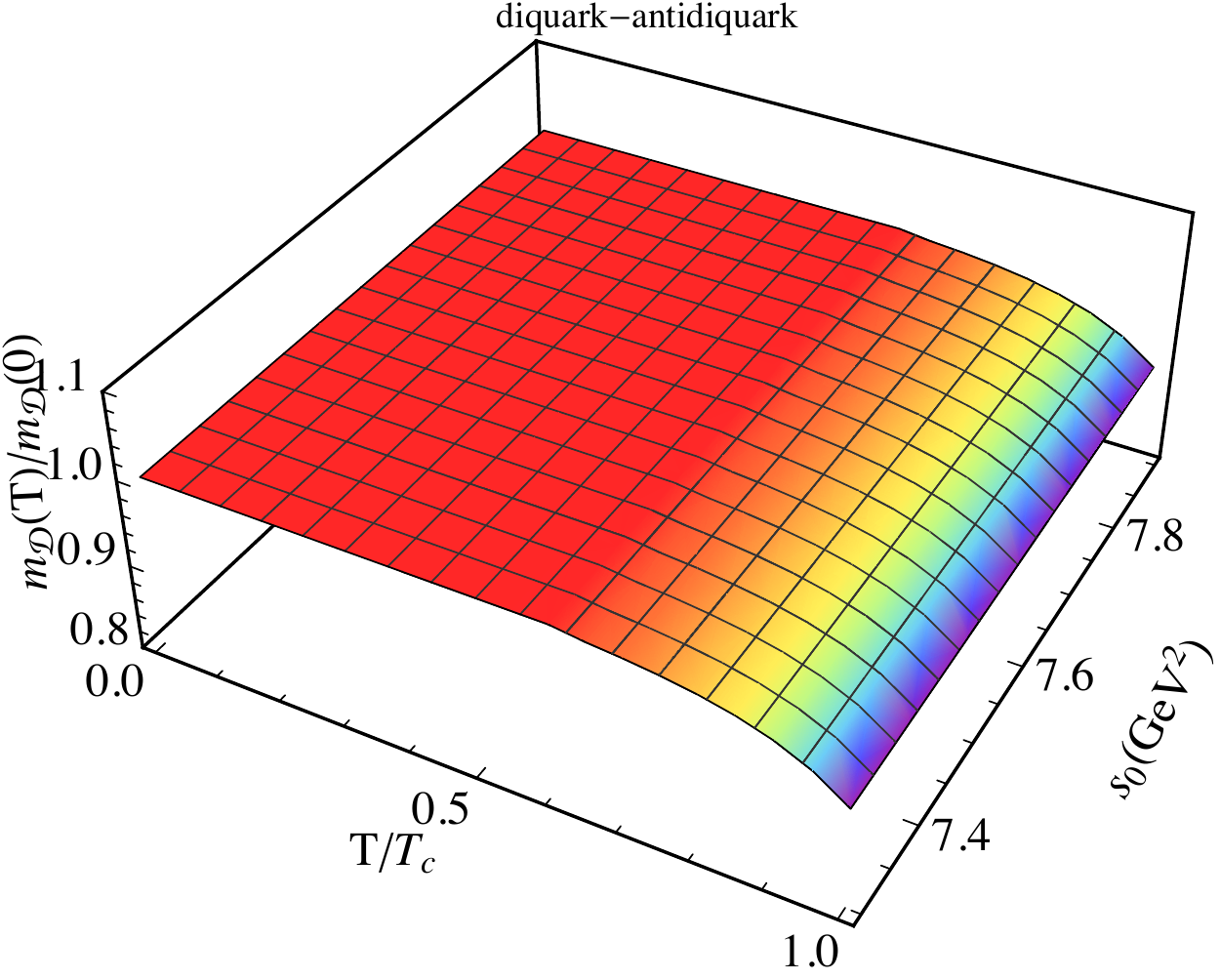}
\end{center}
\caption{The ratio of temperature-dependent mass $m_{\mathcal{D}}(T)$ to zero temperature mass $m_{\mathcal{D}}(0)$ of the $D^*_{s0}(2317)$  state in the molecular and diquark-antidiquark pictures for different fixed values of $s_0(0)$.} \label{fig2}
\end{figure}

\begin{figure}[htbp]
\begin{center}
\includegraphics[width=8.5cm]{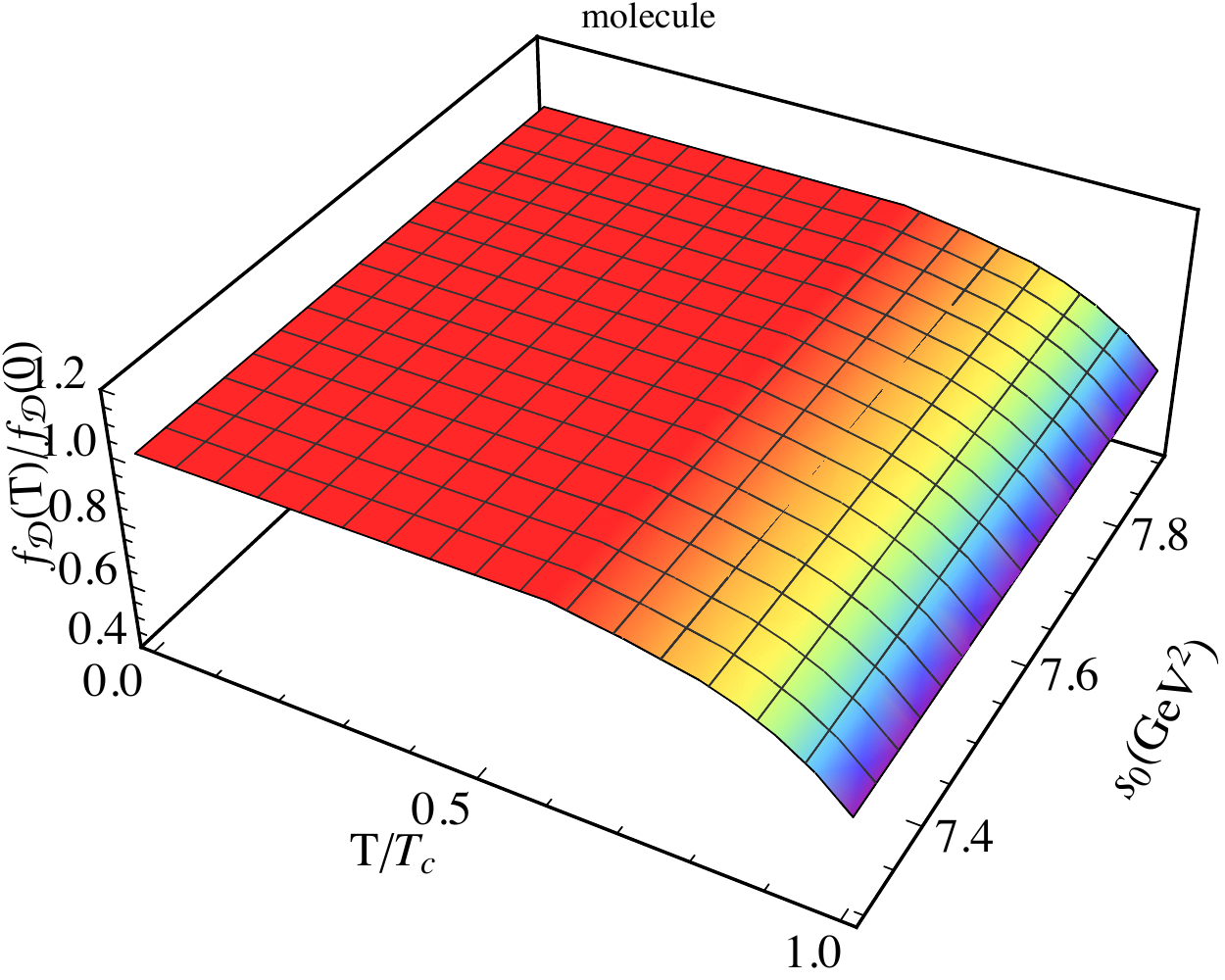}\\[0.7cm]
\includegraphics[width=8.5cm]{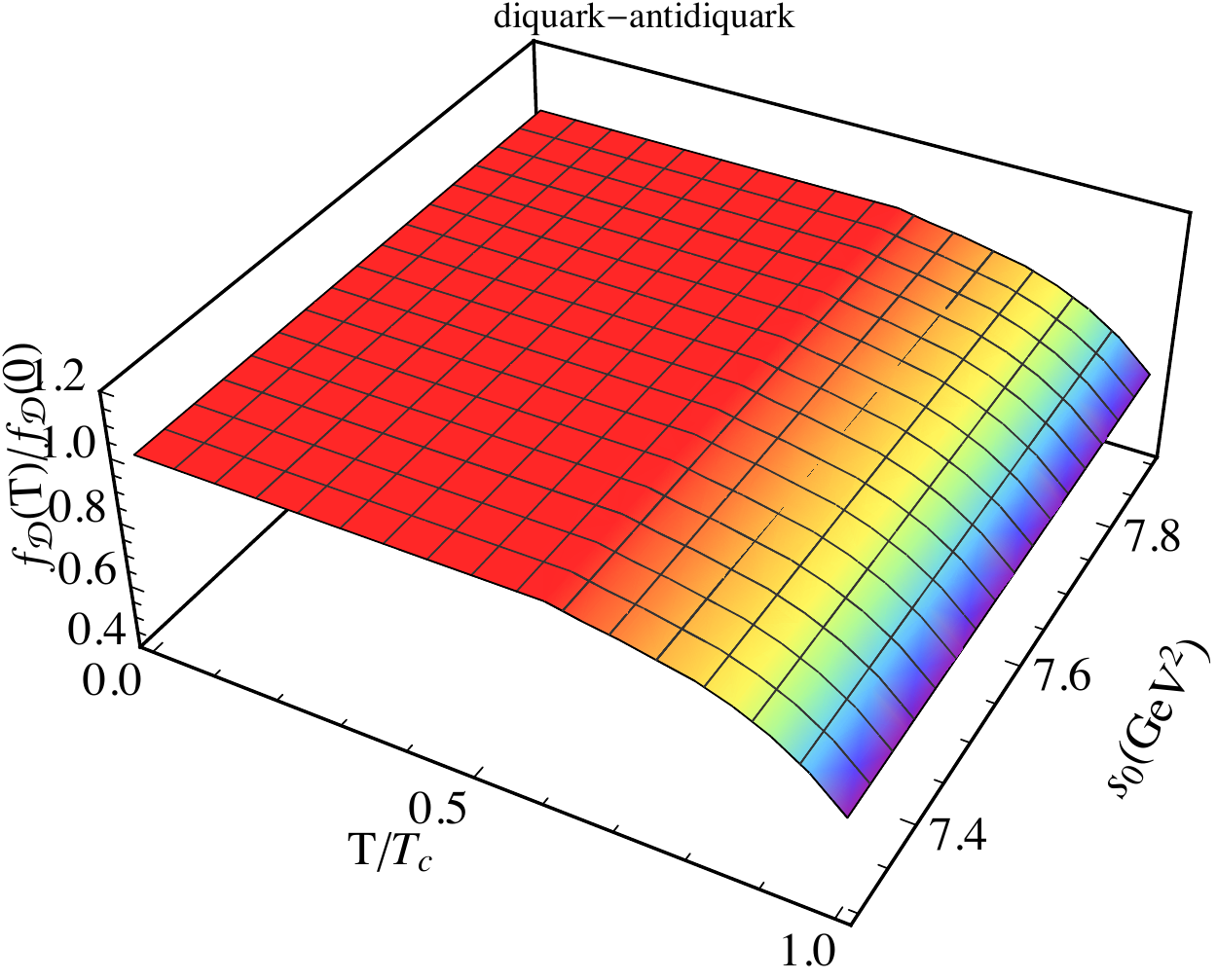}
\end{center}
\caption{The ratio of the temperature-dependent pole residue $f_{\mathcal{D}}(T)$ to zero temperature pole residue $f_{\mathcal{D}}(0)$ of the $D^*_{s0}(2317)$ state in the molecular and diquark-antidiquark pictures for different fixed values of $s_0(0)$.} \label{fig3}
\end{figure}

The last step is to look for changing of the mass and pole
residue of the $D^*_{s0}(2317)$ state in terms of temperature. In
this context the ratio of changing the mass and pole residue graphs are drawn as a
function of the temperature for the molecular and diquark-antidiquark assumptions in Figures \ref{fig2} and \ref{fig3}, respectively. 

As a by product we interchange the $ c $ and $ b $ quarks in the interpolating current and deduce from SU(2) symmetry that the bottom partner of $D^*_{s0}(2317)$ which we name this resonance as $ B_{sJ} $ can possibly between the following mass and pole residue values in both molecular and diquark-antidiquark pictures, respectively:\\
\begin{eqnarray*}\label{eq:BResults}
m_{B_{sJ}}^{mol.}&=&(5140-5458)~\mathrm{MeV},~f_{B_{sJ}}^{mol.}=(33.2-53.3)~\mathrm{keV},\\\\
m_{B_{sJ}}^{di.}&=&(5142-5460)~\mathrm{MeV},~f_{B_{sJ}}^{di.}=(38.3-61.5)~\mathrm{keV}
\end{eqnarray*}\\
which are consistent with the estimations in Ref.~\cite{Albuquerque:2016nlw} within the limits of uncertainties.
%%%%%%%%%%%%%%%%%%%%%%%%%%%%%%%%%%%%%%%%%%%%%%%%%%%%%%%%%%%%%%%%%%
\section{Conclusion}
%%%%%%%%%%%%%%%%%%%%%%%%%%%%%%%%%%%%%%%%%%%%%%%%%%%%%%%%%%%%%%%%%%
Theoretical and experimental disagreement on mass and decay width
of open-charm  meson $D^*_{s0}(2317)$ stimulates several
non-conventional (exotic) interpretation of this state, such as in
the molecular and diquark-antidiquark structures predominantly in the
literature. Its measured mass and width does not match the
predictions from potential-based quark models. Therefore it is so
important to make further studies to finalize this uncertainties.

In this sense we examined the $D^*_{s0}(2317)$ state in the
molecular and diquark-antidiquark pictures by calculating its spectroscopic
parameters using TQCDSR method. We can summarize our results as
follows:

$\bullet$ Our numerical calculations show that mass and pole
residue parameters are practically independent of temperature at
least up to a temperature of $~100$ MeV, but right after that point
they begin to drop by growing temperature.

$\bullet$ Near the critical temperature, the pole residue arrives
to $70\%$ and $69\%$ of its vacuum value in molecular 
and diquark-antidiquark pictures, respectively while the masses decrease by
$9\%$ for both pictures. As for the mass, our result is
in good agreement with Ref.~\cite{Zyla}.

$\bullet$ Our results does not give any definite information as to whether $D^*_{s0}(2317)$ resonance is molecular or diquark-antidiquark structure since they are within the uncertainties of the TQCDSR theory.

$\bullet$ Although there is no scalar particle yet detected in the bottom-strange meson list of the PDG, we predict that it could be found in experiments in the near future.

As a result, the noticeable decline in the values of mass and pole
residue can be conceivable as a manifestation of the QGP phase
transition. The QGP provides a very good ground to understand the
nonperturbative aspects of strong interactions, early universe,
and astrophysical processes such as neutron stars. In short, the
investigation of QGP is quite important to reveal the nature of
both micro and macro scale events in our universe. For this reason
searches related with thermal properties of hadrons and also
 exotic candidates can provide valuable hints and information
for the future experiments such as CMS, LHCb and PANDA.\\
\appendix
\section{The two-point thermal spectral densities}
Here, we present the results of our evaluations for the spectral
density
\begin{eqnarray}
\rho^{\mathrm{QCD}}(s,T)=\rho^{\mathrm{pert.}}(s)+\sum_{k=3}^{6}\rho
_{k}(s,T),  \label{eq:A1}
\end{eqnarray}
essential for our calculations of the mass and pole residue as a
function of the temperature belonging to the $D^*_{s0}(2317)$
resonance via TQCDSR in the diquark-antidiquark picture as an example. $\rho_{k}(s,T)$
denote the nonperturbative contributions to $\rho^{QCD}(s,T)$ and $%
g_s^2=4\pi\alpha_s$. The explicit form for
$\rho^{\mathrm{pert.}}(s)$ and $\rho_{k}(s,T)$ are expressed with
the integrals over the Feynman parameter $z$ as follows:
\begin{eqnarray}\label{eq:rhopert}
	\rho^{pert}(s)&=&\int_{0}^{1} dz\frac{-z^2 \left(m_{c}^2+s \beta\right)^2}{3 \times 2^{11} \pi ^6 \beta^3} \Big(m_{c}^4 z^2-4 m_{c}^3 z (m_{d}\nonumber\\
	&+&m_{u})+4 m_{c}^2 \beta z \big[2 m_{s} (m_{d}+m_{u})+s z \big]-4m_{c}\notag\\ 
	&\times& \beta \big[9 m_{s} (m_{d}^2+m_{u}^2)+s z(m_{d}+m_{u})\big]+s \beta^2 z \notag\\ 
	&\times&\big[20 m_{s} (m_{d}+m_{u})+3 sz \big]\Big)\Theta[L(s,z)]
\end{eqnarray}
\begin{widetext}
\begin{eqnarray}\label{eq:dim3}
\rho^{\langle\bar{q}q\rangle}(s,T)&=&\int_{0}^{1} dz \frac{-z}{2^7 \pi ^4 \beta^2} \Big[\langle\bar{d}d\rangle \Big(m_{c}^5 z+2 m_{c}^4 \beta z (m_{d}-m_{s})+2
m_{c}^3 \beta (-m_{d}^2+4 m_{d} m_{s}+s z)
+2 m_{c}^2 \beta^2\big[2 m_{d}^2 m_{s}\nonumber\\
&+&3 s z (m_{d}-m_{s})\big]+m_{c} s \beta^2 (-2 m_{d}^2+8 m_{d} m_{s}+s z)+2 s \beta^3 \big[3 m_{d}^2
m_{s}+2 s z (m_{d}-m_{s})\big]\Big)+m_{c}^5 \langle\bar{u}u\rangle z \nonumber\\ 
&-& 2 m_{c}^4
\beta z \Big(\langle\bar{s}s\rangle (m_{d}-m_{s}+m_{u})+\langle\bar{u}u\rangle (m_{s}-m_{u})\Big)+
2m_{c}^3 \beta \Big(2 m_{d}^2 \langle\bar{s}s\rangle-m_{d} m_{s} \langle\bar{s}s\rangle+m_{u}
\big[-m_{s} \langle\bar{s}s\rangle\nonumber\\
&+&4 m_{s} \langle\bar{u}u\rangle+2 m_{u} \langle\bar{s}s\rangle-m_{u} \langle\bar{u}u\rangle \big]+s
\langle\bar{u}u\rangle z\Big)
+2 m_{c}^2 \beta^2 \Big(2 m_{s} m_{u}^2 \langle\bar{u}u\rangle-3 s z
\big[\langle\bar{s}s\rangle (m_{d}-m_{s}+m_{u})\nonumber\\
&+&\langle\bar{u}u\rangle
(m_{s}-m_{u})\big]\Big)+m_{c} s \beta^2 \Big(4 m_{d}^2 \langle\bar{s}s\rangle-2 m_{d}
m_{s} \langle\bar{s}s\rangle-2 m_{u} \big[m_{s} \big(\langle\bar{s}s\rangle-4 \langle\bar{u}u\rangle \big) +m_{u} \big(\langle\bar{u}u\rangle-2
\langle\bar{s}s\rangle \big)\big]\nonumber\\
&+& s \langle\bar{u}u\rangle z\Big)-
2 s \beta^3 \Big(2 s z \big[\langle\bar{s}s\rangle
\big(m_{d}-m_{s}+m_{u}\big)+\langle\bar{u}u\rangle \big(m_{s}-m_{u}\big)\big]-3 m_{s} m_{u}^2
\langle\bar{u}u\rangle\Big)\Big]\Theta[L(s,z)]
\end{eqnarray}
\begin{eqnarray}\label{eq:dim4}
\rho^{G^2+\langle \theta_{00}\rangle}(s,T)&=&\int_{0}^{1} dz \bigg[\frac{1}{3^2 \times 2^{12} \beta^3 \pi^6}z \Big(-16 \beta^2 \Big[6 s z m_c^2 \Big(8 \beta \pi^2 \langle u^{\mu} \theta^f_{\mu\nu} u^{\nu} \rangle \
(-3 + 16 z) +  g_s^2\langle u^{\mu} \theta^g_{\mu\nu} u^{\nu} \rangle [3 + 4 z \notag\\ 
&\times& (-3 + 2 z)] \Big)  + 6 \beta z m_c^4 \Big(16 \pi^2 \langle u^{\mu} \theta^f_{\nu} u^{\nu} \rangle + g_s^2 \langle u^{\mu} \theta^g_{\mu\nu} u^{\nu} \rangle  \Big)  - 6 \beta s (-1 + 3 z) \Big[16 \pi^2 \langle u^{\mu} \theta^f_{\mu\nu} u^{\nu} \rangle \notag\\ 
&+& \langle u^{\mu} \theta^g_{\mu\nu} u^{\nu} \rangle g_s^2 \Big]  - 6 \beta m_c^3(m_d +m_u)\Big(16 \pi^2 \langle u^{\mu} \theta^f_{\mu\nu} u^{\nu} \rangle+\langle u^{\mu} \theta^g_{\mu\nu} u^{\nu} \rangle  g_s^2\Big)   + \beta s \Big(2 s z \big[8 \beta \pi^2\notag\\ 
&\times&  \langle u^{\mu} \theta^f_{\mu\nu} u^{\nu} \rangle (-3 + 50 z) + g_s^2 \langle u^{\mu} \theta^g_{\mu\nu} u^{\nu} \rangle[6 + z (-34 + 25 z)] \big] - 9 m_d m_s z g_s^2\big[-16 \beta \pi^2 \langle u^{\mu} \theta^f_{\mu\nu} u^{\nu} \rangle \notag\\ 
&+&
\langle u^{\mu} \theta^g_{\mu\nu} u^{\nu} \rangle \big] - 
9 m_s m_u \big[-16 \beta \pi^2 \langle u^{\mu} \theta^f_{\mu\nu} u^{\nu} \rangle 
+ z g_s^2 \langle u^{\mu} \theta^g_{\mu\nu} u^{\nu} \rangle \big] \Big)\Big] + 
G^2 g_s^2 \Big[-2 (6 - 5 z)^2z m_c^4\notag\\
&+& 2 \beta^2 s m_c (m_d + m_u) \big[-36 + z (36 + z)\big] + 
72 \beta^2 m_c^3 (m_d+ m_u) + 
12 \beta^3 s z \Big(4 s (3 - 2 z) + 9 m_s \notag\\ 
&\times& (m_d +m_u)\Big) + 
\beta z m_c^2 \Big[-147 s z^2 - 72 \Big(3 s + m_s (m_d + m_u)\Big) + 
4 z \Big(90 s + 17 m_s (m_d + m_u)\Big)\Big]\Big]\Big)\notag\\ 
&\times& \Theta[L(s,z)]
\end{eqnarray}
\begin{eqnarray}\label{eq:dim5}
\rho^{\langle qG q \rangle}(s,T)&=&\int_{0}^{1} dz \Big[\frac{m_0^2 }{3\times 2^7 \beta \pi^4}\Big[3 \langle\bar{u}u\rangle z m_c^3 + 
\langle\bar{d}d\rangle \Big(3 z m_c^3 + 2 \beta z m_c^2 (2 m_d - 3 m_s) + 
\beta m_c (3 s z - m_d^2 + 6 m_d m_s) + 
\beta^2 \notag\\ 
&\times& (6 s z m_d - 9 s z m_s + 2 m_d^2 m_s)\Big) - 
2 \beta z m_c^2 \Big(3 \langle\bar{s}s\rangle m_d - 2 \langle\bar{s}s\rangle m_s + 3 \langle\bar{u}u\rangle m_s + 3 \langle\bar{s}s\rangle m_u - 
2 \langle\bar{u}u\rangle m_u\Big) + 
\beta^2 \notag\\ 
&\times&\Big(2\langle\bar{u}u\rangle m_s m_u^2 - 
3 s z \big[3 \langle\bar{s}s\rangle m_d - 2 \langle\bar{s}s\rangle m_s + 3 \langle\bar{u}u\rangle m_s + 3 \langle\bar{s}s\rangle m_u - 
2 \langle\bar{u}u\rangle m_u\big]\Big) + 
\beta m_c \Big(3 s \langle\bar{u}u\rangle z\notag\\ 
&+& 3 \langle\bar{s}s\rangle m_d^2 - \langle\bar{s}s\rangle m_d m_s + 
m_u \big[-\langle\bar{s}s\rangle m_s + 6 \langle\bar{u}u\rangle m_s + 3 \langle\bar{s}s\rangle m_u - \langle\bar{u}u\rangle m_u \big]\Big)\Big]\Theta[L(s,z)],
\end{eqnarray}
\begin{eqnarray}\label{eq:dim6}
\rho^{\langle\bar{q}q\rangle^2}(s,T)&=&\int_{0}^{1} dz \Big[\frac{1}{9^2\times 2^5 \pi^4} \Big[-54 \langle\bar{d}d\rangle  \pi^2  \langle\bar{s}s\rangle \Big(4 z m_c^2 + m_c (-4 m_d + m_s) + 
2 \beta  \big[2 s z + m_d (-m_d + m_s)\big]\Big) + 
\langle\bar{d}d\rangle^2 \notag\\ 
&\times& 
\Big(-54 \pi^2 (m_c + \beta m_d) (m_d - 2 m_s) + 
g_s^2 \Big[-4 z m_c^2 + m_c m_d - 2  \beta (2 s z + m_d m_s)\Big]\Big) + 
g_s^2 \Big(-4 ( \langle\bar{s}s\rangle^2  \notag\\ 
&+&  \langle\bar{u}u\rangle ^2) z m_c^2 +  \langle\bar{u}u\rangle^2 m_c m_u + 
 \langle\bar{s}s\rangle^2 m_c (m_d + m_u) - 
2  \beta \Big[2 s ( \langle\bar{s}s\rangle^2 +  \langle\bar{u}u\rangle^2) 
z +  \langle\bar{u}u\rangle^2 m_s m_u \Big]\Big) - 
54 \pi^2   \notag\\ 
&\times& \langle\bar{u}u\rangle \Big(4  \langle\bar{s}s\rangle zm_c^2 + m_c \Big[m_s \big( \langle\bar{s}s\rangle - 2  \langle\bar{u}u\rangle \big)  + m_u (-4  \langle\bar{s}s\rangle +  \langle\bar{u}u\rangle) \Big] + 
 \beta \Big[4 s  \langle\bar{s}s\rangle z + m_u [2 ( \langle\bar{s}s\rangle -  \langle\bar{u}u\rangle)  \notag\\ 
 &\times& m_s+ (-2   \langle\bar{s}s\rangle 
  +\langle\bar{u}u\rangle)m_u ]\big]\Big)\Big]\Theta[L(s,z)] \Big],
\end{eqnarray}
\begin{eqnarray}\label{eq:nonpert}
\mathcal{\widehat{B}}\widetilde{\Gamma}(q^2,T)&=&\bigg[\dfrac{1}{3\times 2^5 \pi^2}  m_c m_s \langle\bar{s}s\rangle  \Big(m_d^2\langle\bar{d}d\rangle  + m_u^2\langle\bar{u}u\rangle  \Big) ~e^{-m_c^2/M^2}\bigg]\notag\\
&-&\int_{0}^{1} dz \bigg[\dfrac{z^2}{3 \times 2^{11} \beta^2 \pi^6} G^2  g_s^2 m_c^3 m_s (m_d^2 + m_u^2)~e^{-m_c^2/{M^2\beta}} \bigg]~~~~~~~~~~~~~~~~~~~~~~~~~~~~~~~~~~~~~~~~~~~~~~~~~~~~~~~~~~
\end{eqnarray}
where $\Theta $ denotes the unit step function,  $L(s,z)=sz(1-z)-zm_{c}^2 $ and $ \beta=z-1 $.
\end{widetext}

\end{document}